\documentclass[preprint,12pt,a4paper]{elsarticle}

\usepackage{amssymb}

\begin{document}

\begin{frontmatter}

\title{Illumination effect by visible continuous-wave laser on bulk 40 K-superconductors}

\author{Jiro Kitagawa$^1$}

\address{$^1$ Department of Electrical Engineering, Faculty of Engineering, Fukuoka Institute of Technology, 3-30-1 Wajiro-higashi, Higashi-ku, Fukuoka 811-0295, Japan}
\ead{j-kitagawa@fit.ac.jp}

\author{Tatsuya Yoshika$^1$}

\address{$^1$ Department of Electrical Engineering, Faculty of Engineering, Fukuoka Institute of Technology, 3-30-1 Wajiro-higashi, Higashi-ku, Fukuoka 811-0295, Japan}

\author{Nobuhiro Gonda$^1$}

\address{$^1$ Department of Electrical Engineering, Faculty of Engineering, Fukuoka Institute of Technology, 3-30-1 Wajiro-higashi, Higashi-ku, Fukuoka 811-0295, Japan}

\begin{abstract}
Visible CW (continuous wave) -laser heating effects on the bulk superconductors CeFeAsO$_{0.65}$F$_{0.35}$ and MgB$_{2}$ with 1.5 mm thickness have been investigated by measuring the temperature dependence of electrical resistivity. Each compound shows a critical-temperature $T_{c}$ reduction with increasing fluence rate. At the normal state, a parallel circuit model based on the Fourier's law can well reproduce the temperature dependence of electrical resistivity of illuminated sample. On the other hand, the predicted temperature-rise due to the laser heating in the superconducting state is much smaller than the observed $T_{c}$-reduction. A temperature gradient of a few K across the sample thickness easily triggers the destruction of bulk superconductivity. Furthermore we have found a slight $T_{c}$-enhancement in CeFeAsO$_{0.65}$F$_{0.35}$ after a rather high fluence-rate irradiation.
\end{abstract}

\begin{keyword}
Fe-based superconductor; MgB$_{2}$; Illumination effect; Laser heating
\end{keyword}

\end{frontmatter}

\clearpage

\section{Introduction}
Light-matter interactions are widely studied, because the light is one of attractive external-field sources.
For example, light can induce metallic states or opto-mechanical effects in semiconductors\cite{Katsumoto:JPSJ1987,Kundys:PRB2012}.
In addition dielectric property of material can be tuned by light\cite{Takesada:JPSJ2003,Hasegawa:JPSJ2003}.
Furthermore so-called strongly-correlated electron systems offer good opportunities to study light-matter interactions.
For $d$-electron systems, photoinduced metallic states in Mott insulators\cite{Miyano:PRL1997,Iwai:PRL2003,Kitagawa:JPCM2007}, photoinduced magnetization\cite{Sato:Science1996}, and a photoinduced change in spin configuration\cite{Ogawa:PRL2000} et al. have been reported.
The illumination effects on $f$-electron systems have been extensively studied for Eu chalcogenides\cite{Umehara:PRB1995,Liu:PRL2012}.
The photocarrier doping in semiconductors and the optical control of the Kondo effect are recently reported\cite{Kitagawa:JPSJ2013,Kozasa:MRX2014,Kitagawa:PRB2016}. 

Focusing on optical responses of superconductors, which belong to another class of attractive platform to investigate the interaction between many-electron systems and light, there exists a long history of studies, usually employing thin film samples.
Testardi demonstrated the destruction of superconductivity in Pb films\cite{Testardi:PRB1971}.
When he used a pulse visible-laser with 514 nm wavelength at the fluence rate of 3 Wcm$^{-2}$, the reduction of superconducting critical temperature $T_{c}$ was 3.2 K, in which the temperature rise by the laser heating was 0.45 K.
The main origin of $T_{c}$ reduction is a Cooper pair breaking by photons with the energy larger than $T_{c}$. 

The discoveries of high-$T_{c}$ superconductors have accelerated efforts to control the carrier number through photocarrier doping. 
In Y-Ba-Cu-O superconductors, $T_{c}$ increases under the illumination of a laser\cite{Nieva:APL1992,Osada:PRB2005}, which is caused by a carrier doping through a microscopic structural-change.
There are many studies of time-resolved optical pump-probe measurements, observing a pico second recovery process of broken Cooper-pairs\cite{Kavanov:PRB1999,Demsar:PRL2003}.
More recently light-induced superconductivities are demonstrated by tuning the wavelength of light in a cuprate and a fullerene\cite{Fausti:Science2011,Mitrano:Nature2016}.

In this paper we have studied the illumination effect on bulk superconductors. 
The optical penetration depth is much smaller than the sample thickness.
Therefore the contribution of Cooper-pair breaking would be neglected.
If a continuous wave (CW) laser is used, we expect a nearly equilibrium state and can discuss an accurate thermal effect by laser heating on a bulk superconductor, which has been overlooked.
In this study, we present the detailed studies of CW-laser heating by employing CeFeAsO$_{0.65}$F$_{0.35}$ and MgB$_{2}$ 40 K-superconductors\cite{Kamihara:JACS2008,Chen:PRL2008,Nagamatsu:Nature2001}.
In addition we have found a laser annealing effect in the Fe-based superconductor.

\section{Materials and methods}
All samples were synthesized by the solid state reaction technique.
For CeFeAsO$_{0.65}$F$_{0.35}$, the precursor Fe$_{2}$As (Ce$_{1.117}$As) was prepared by reacting the constituent element powders with the molar ratio in an evacuated quartz tube, that was heated at 900 $^{\circ}$C.
The powdered Fe$_{2}$As and Ce$_{1.117}$As were homogeneously mixed with CeO$_{2}$ and CeF$_{3}$ powders with the molar ratio Fe$_{2}$As:Ce$_{1.117}$As:CeO$_{2}$:CeF$_{3}$$=$1:1:0.65:0.233.
The mixture was pressed into a pellet, which was heated in an evacuated quartz tube to 1050 $^{\circ}$C.
For MgB$_{2}$, Mg and B powders were homogeneously mixed.
The pressed pellet was placed on an Ta sheet in a sealed quartz tube partially filled by Ar gas.
The tube was heated to 900 $^{\circ}$C and cooled down to 600 $^{\circ}$C at the rate of 30 $^{\circ}$C/h, and subsequently quenched in water.
All products were evaluated using a powder X-ray diffractometer with Cu-K$\alpha$ radiation.
The powder X-ray diffraction patterns show almost single phases with small amount of impurity phases.

The temperature dependence of the electrical resistivity $\rho(T)$ between 20 and 300 K under illumination was measured by the conventional DC four-probe method using a closed-cycle He gas cryostat.
The optical source was a CW laser diode (LD) with photon energy of 1.85 eV, which was sufficient for destroying Cooper pairs of studied superconductors. 
A parallelepiped sample with dimensions of $\sim$1.5$\times$1.5$\times$10 mm$^{3}$ was cut from a pellet of each sample, and the distance between the voltage electrodes was 2 to 3 mm. 
The sample was glued using varnish onto a thin Cu plate covered with a cigarette paper. 
The Cu plate was then placed on the cooling stage in the cryostat using Apiezon N grease as an adhesive that provides good thermal contact.
The temperature was measured by a thermocouple, thermally anchored to the cooling stage underneath the sample.
All electrodes were covered by thin phosphor-bronze plates to reduce any extrinsic photovoltaic effects.
The light beam from the LD was focused onto the area between the voltage electrodes. 

\section{Results and discussion}
Figure 1(a) shows $\rho(T)$ of CeFeAsO$_{0.65}$F$_{0.35}$ under the fluence rates denoted in the figure. 
The sample in the dark state undergoes superconductivity below $T_{c}$ of 39.5 K, determined by the midpoint of resistance drop.
Increasing fluence rate depresses the superconductivity and $T_{c}$ is not observed down to 25 K at 3.5 Wcm$^{-2}$.
$\rho(T)$ at higher than 75 K does not change significantly by the illumination.
MgB$_{2}$ shows $T_{c}$ of 39 K at the dark state (see Fig.\ 1(b)). 
By increasing the fluence rate, $T_{c}$ is gradually reduced.
We have observed the similar trend of fluence-rate dependence of $T_{c}$. 
Thus the $T_{c}$ reduction under illumination is not characteristic only for the Fe-based superconductor.

\begin{figure}[hbtp]
\begin{center}
\includegraphics[width=0.8\linewidth]{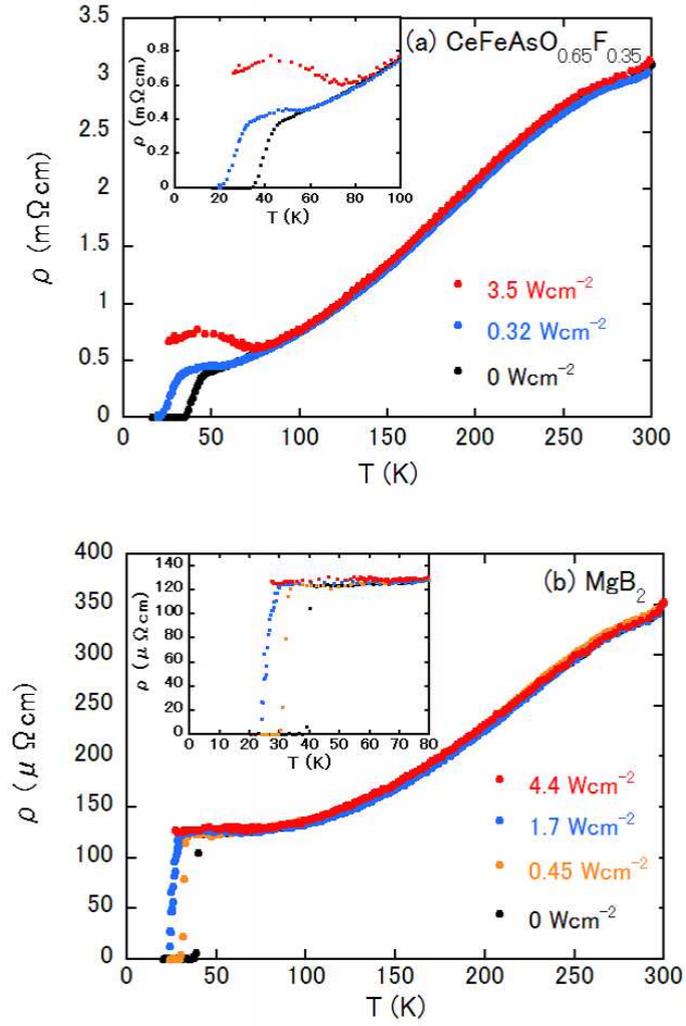}
\caption{(a)Temperature dependence of $\rho$ of CeFeAsO$_{0.65}$F$_{0.35}$ under fluence rate of 0, 0.32 and 3.5 Wcm$^{-2}$. The inset is the expanded view at low temperatures. (b)Temperature dependence of $\rho$ of MgB$_{2}$ under fluence rate of 0, 0.45, 1.7 and 4.4 Wcm$^{-2}$. The inset is the expanded view at low temperatures.}
\end{center}
\end{figure}

The optical penetration depths of CeFeAsO$_{0.65}$F$_{0.35}$ and MgB$_{2}$ at 1.85 eV are estimated to be 45 nm and 21 nm, respectively\cite{penetration depth}, using permittivity (and optical conductivity for MgB$_{2}$) data\cite{Boris:PRL2009,Guritanu:PRB2006}.
Because each optical penetration depth is much smaller than the sample thickness $d_{s}$ 1.5 mm, it is not conceivable that the Cooper-pair breaking by laser light occurs in the whole sample volume.
Hereafter the thermal effect due to heating, which is evaluated by a model based on the Fourier's law, is discussed.

If all of the light power with fluence rate $F$ is transformed into heat in a sample, according to the Fourier's law, the temperature rise $\Delta T$ across the sample thickness under an equilibrium condition can be expressed by

\begin{equation}
\Delta T=\frac{F d_{s}}{\kappa}, 
\label{equ:temprise}
\end{equation}
where $\kappa$ is the thermal conductivity of sample.
The conditions of application of Eq.\ (1) are $d_{s}$ $\gg$ phonon mean free path and macroscopic measurement time $\gg$ phonon relaxation time.
$d_{s}$ of each sample is 1.5 mm, which is much longer than the phonon mean free path at most of the order of the lattice constant\cite{Naidyuk:PRL2003}.
The macroscopic measurement time of the order of second in this study, is also much longer than the phonon relaxation time, which is usually in pico second order\cite{Demsar:PRL2003,Mertelj:JSNM2009}.
These facts satisfy the assumptions of Fourier's law.
Since the optical penetration depth is much smaller than $d_{s}$, an inhomogeneous temperature gradient in the illuminated volume can be neglected.
Therefore the temperature gradient along the sample thickness would be approximately constant, which is the additional important factor for the application of Fourier's law.

\begin{figure}[hbtp]
\begin{center}
\includegraphics[width=0.6\linewidth]{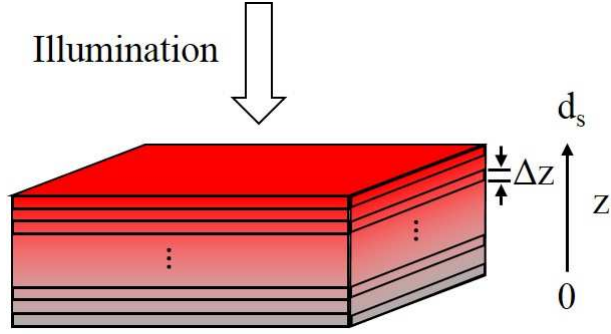}
\caption{Schematic of photoexcited sample in the equilibrium state. The temperature gradient along the sample depth $z$ is expressed by color gradation. The thin platelet with the black frame is a discretized layer, in which the temperature is assumed to be constant.}
\end{center}
\end{figure}

The sample is discretized along the sample depth $z$ by the platelet as thin as possible with the thickness $\Delta z$ (see Fig.\ 2).
Assuming the constant temperature-gradient between the illuminated top surface and the sample bottom with the temperature $T$ measured by the thermocouple, $\rho$(T) at the $i$-th layer from the bottom is expressed by $\rho (T+\Delta T\frac{\Delta z}{d_{s}}i$).
Regarding the sample as the parallel circuit of discretized layers, $\rho$ of whole volume is expressed as follows;
\begin{equation}
\rho=\frac{d_{s}}{\sum_{i}\frac{\Delta z}{\rho(T+\Delta T\frac{\Delta z}{d_{s}}i)}}. 
\label{equ:rho}
\end{equation}

By employing the reported $\kappa$-data\cite{McGuire:NJP2009,Bauer:JPCM2001,Schneider:PhysicaC2001} of CeFeAsO and MgB$_{2}$, the temperature dependences of calculated $\rho$ in the normal state under selected fluence-rates are presented in Figs.\ 3(a) and 3(b) by solid lines.
In Fig.\ 3(a), slightly larger calculated-$\rho$ above 75 K compared to the experimental $\rho$, especially at 3.5 Wcm$^{-2}$, is due to the overestimated $\Delta T$, which is caused by the employed rather low $\kappa$ and/or the neglect of thermal dissipation path in the discretized layer.
The predicted $\rho$ of MgB$_{2}$ is in good agreement with experimental one above approximately 70 K (see Fig.\ 3(b)).
Figures 3(a) and 3(b) suggest that the parallel circuit model with the Fourier's law can semi-quantitatively explain the laser heating of $\rho(T)$ in the normal state.

\begin{figure}[hbtp]
\begin{center}
\includegraphics[width=0.8\linewidth]{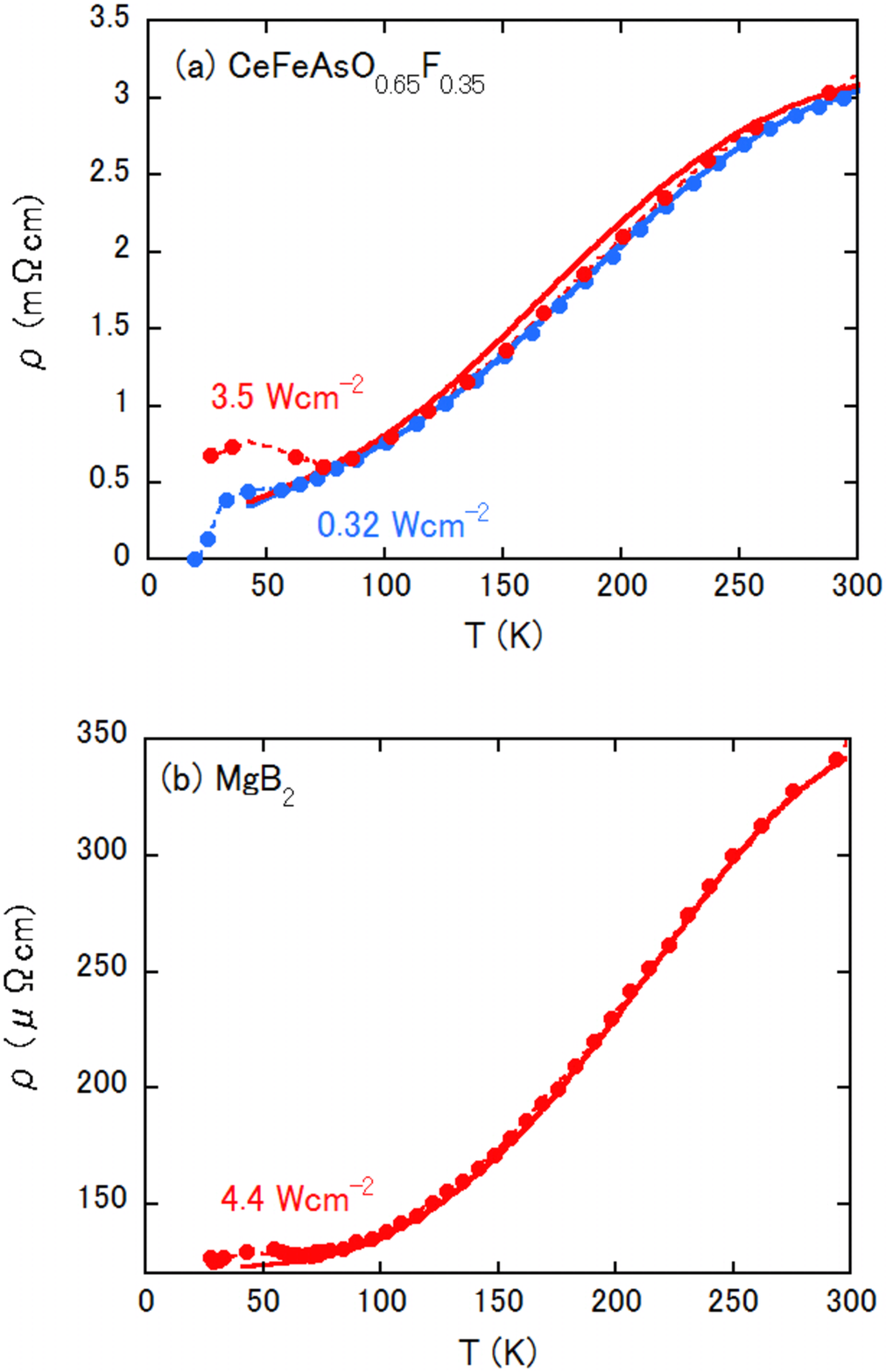}
\caption{(a)Temperature dependence of $\rho$ of CeFeAsO$_{0.65}$F$_{0.35}$ under fluence rate of 0.32 and 3.5 Wcm$^{-2}$ (filled circles with broken lines). The solid lines are the normal-state $\rho$ under illumination (blue: 0.32 Wcm$^{-2}$, red: 3.5 Wcm$^{-2}$), calculated by the model based on the Fourier's law. (b)Temperature dependence of $\rho$ of MgB$_{2}$ under fluence rate of 4.4 Wcm$^{-2}$ (filled circles with broken line). The solid line is the normal-state $\rho$ under the illumination, calculated by the model based on the Fourier's law.}
\end{center}
\end{figure}

Figure 4 displays the temperature dependence of $\Delta T_{ave}$, determined by averaging $\Delta T$ over the whole sample, at low temperatures for each compound under a higher fluence rate.
If the superconducting state also follows the Fourier's law, $T_{c}$ reduction can be evaluated using $\Delta T_{ave}$.
The $T_{c}$ reductions are 3.5 K (at 3.5 Wcm$^{-2}$) for CeFeAsO$_{0.65}$F$_{0.35}$ and 2.4 K (at 4.4 Wcm$^{-2}$) for MgB$_{2}$, respectively.
The actual $T_{c}$ reduction for each compound is above approximately 15 K, largely exceeding the estimated temperature-rise due to laser heating.
Therefore the Fourier's law would not be a dominant factor, leading to the $T_{c}$-reduction under illumination.

In the report\cite{Testardi:PRB1971} of Pb thin film with 27.5 nm thickness, the temperature rise due to laser heating at 3 Wcm$^{-2}$ is 0.45 K, which is smaller than the $T_{c}$ reduction of 3.2 K.
With increasing film-thickness up to 200 nm, only the laser heating of 0.45 K is obtained.
However, for bulk sample in this study, $T_{c}$ reduction again largely exceeds the predicted temperature-rise due to laser heating, as in Pb thin film with 27.5 nm thickness.
Since the Cooper-pair breaking as in thin film is not operative in the bulk sample, a model other than the Fourier's law is responsible for the observed $T_{c}$-reductions in CeFeAsO$_{0.65}$F$_{0.35}$ and MgB$_{2}$.
Our findings strongly suggest that the temperature gradient of the order of a few K across the sample thickness triggers the destruction of bulk superconductivity.
The temperature gradient easily produced by LD might lead to a thermal-instability phenomenon similar to the quench.

\begin{figure}[hbtp]
\begin{center}
\includegraphics[width=0.6\linewidth]{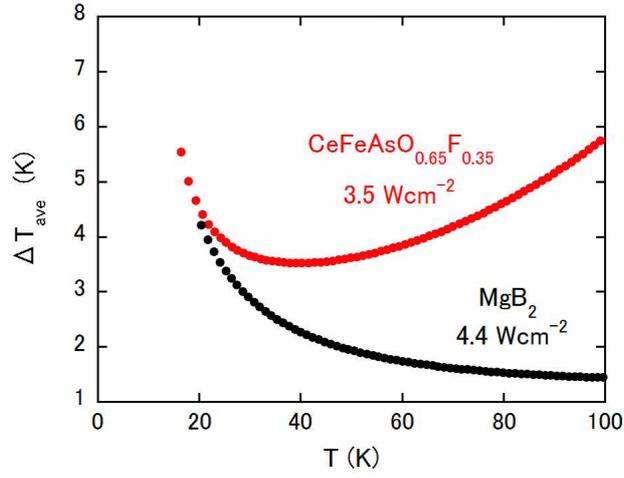}
\caption{Temperature dependence of temperature-rise averaged over the whole sample under laser heating, which is calculated according to the Fourier's law, for CeFeAsO$_{0.65}$F$_{0.35}$ (at 3.5 Wcm$^{-2}$) and MgB$_{2}$ (at 4.4 Wcm$^{-2}$).}
\end{center}
\end{figure}

\begin{figure}[hbtp]
\begin{center}
\includegraphics[width=0.6\linewidth]{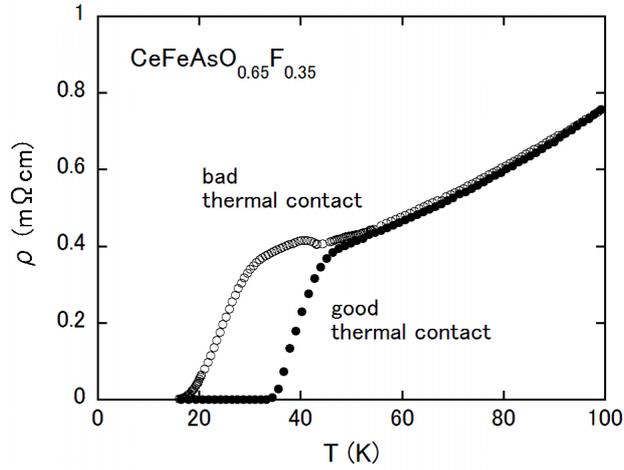}
\caption{Temperature dependence of $\rho$ of CeFeAsO$_{0.65}$F$_{0.35}$ with good (filled circles) and bad (open circles) thermal contacts between sample and Cu plate.}
\end{center}
\end{figure}

To get more deep insight of the heating effect on bulk superconductivity, $\rho(T)$ of CeFeAsO$_{0.65}$F$_{0.35}$ with a bad thermal contact between the sample and the Cu plate was measured as shown in Fig.\ 5 (see the open circles).
The bad thermal contact was realized by a loose varnish-gluing.
The filled circle data showing $T_{c}$ of 39.5 K is taken at the condition of a good thermal contact.
When the thermal contact is weakened, $T_{c}$ is reduced to 25 K, while $\rho(T)$ at the normal state is not largely altered.
Even in the case of bad thermal contact, the actual averaged sample-temperature at $T_{c}$ of 25 K should be 39 K.
This supports that the effect of temperature-gradient across the sample in the superconducting state is different from that in the normal state.

In this study on CeFeAsO$_{0.65}$F$_{0.35}$, we have found that $T_{c}$ is slightly enhanced after a high fluence-rate irradiation.
Figure 6 shows the comparison of $\rho(T)$ of CeFeAsO$_{0.65}$F$_{0.35}$ at the dark state between before starting illumination experiments and after 6.1 Wcm$^{-2}$ irradiation.
The latter state shows $T_{c}$ of 44 K, and $\rho(T)$ smaller than that of the former state, suggesting a possible laser-annealing effect.

\begin{figure}[hbtp]
\begin{center}
\includegraphics[width=0.6\linewidth]{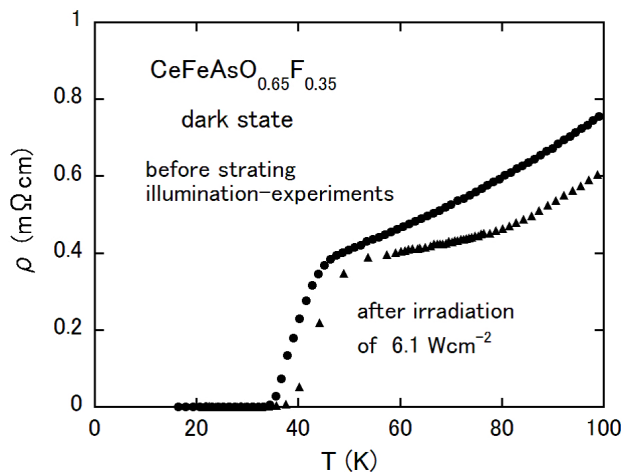}
\caption{Comparison of $\rho$ of CeFeAsO$_{0.65}$F$_{0.35}$ at dark state between before starting illumination-experiments (circles) and after high fluence-rate irradiation of 6.1 Wcm$^{-2}$ (triangles).}
\end{center}
\end{figure}

\section{Summary}
The thermal effect by laser heating in bulk superconductors has been overlooked.
In this paper, detailed studies on the illumination effects of bulk CeFeAsO$_{0.65}$F$_{0.35}$ and MgB$_{2}$ 40 K-superconductors using CW LD have been carried out.
The CW LD allows the observation of equilibrium photoexcited state.
The illumination effect on the normal-state resistivity can be quantitatively explained by the parallel circuit model with the Fourier's law.
However, the estimated temperature-rise due to the laser heating is much smaller than the observed $T_{c}$-reduction.
A model other than the Fourier's law would be operative in the thermal effect on superconducting state. 
In addition $T_{c}$ of CeFeAsO$_{0.65}$F$_{0.35}$ is slightly enhanced after the high fluence-rate irradiation.

\section*{Acknowledgement}
J.K. is grateful for the support provided by the Comprehensive Research Organization of Fukuoka Institute of Technology.

\end{document}